\documentclass[12pt]{iopart}
\usepackage{graphicx}
\begin{document}

\title[Artificial Neural Networks and BDT]{Use of Machine Learning Technique to maximize the signal over background for $H \rightarrow \tau \tau$}

\author{Kanhaiya Gupta}

\address{Physikalisches Institut, Universität Bonn, Nussallee 12, 53115 Bonn, Germany}
\ead{kanhaiya.gupta@uni-bonn.de}
\vspace{10pt}
\begin{indented}
\item[]July 2021
\end{indented}

\begin{abstract}
In recent years, artificial neural networks (ANNs) have won numerous contests in pattern recognition and machine learning. ANNS have been applied to problems ranging from speech recognition to prediction of protein secondary structure, classification of cancers and gene prediction. Here, we intend to maximize the chances of finding the Higgs boson decays to two $\tau$ leptons in the pseudo dataset using a Machine Learning technique to classify the recorded events as signal or background.
\end{abstract}

%
% Uncomment for keywords
%\vspace{2pc}
\noindent{\it Keywords}: Neural Networks, Tau lepton, ROC Curve,  Boosted Decision Tree
%
% Uncomment for Submitted to journal title message
%\submitto{\JPA}
%
% Uncomment if a separate title page is required
%\maketitle
% 
% For two-column output uncomment the next line and choose [10pt] rather than [12pt] in the \documentclass declaration
%\ioptwocol
%

\section{Introduction}
A standard neural network (NN) consists of many simple, connected processors called neurons, each producing a sequence of real-valued activations. Neural-network algorithms for machine learning are inspired by the architecture and the dynamics of networks of neurons in the brain. The algorithms use highly idealised neuron models. Nevertheless, the fundamental principle is the same: artificial neural networks learn by changing the connections between their neurons. Such networks can perform a multitude of information-processing tasks. Input neurons get activated through sensors perceiving the environment, other neurons get activated through weighted connections from previously active neurons.

Neural networks can for instance learn to recognise structures in a set of “training” data and, to some extent, generalise what they have learnt. A training set contains a list of input patterns together with a list of corresponding labels, or target values, that encode the properties of the input patterns the network is supposed to learn. Artificial neural networks can be trained to classify such data very accurately by adjusting the connection strengths between their neurons, and can learn to generalise the result to other data sets provided that the new data is not too different from the training data. A prime example for a problem of this type is object recognition in images, for instance in the sequence of camera images taken by a self-driving car. Recent interest in machine learning with neural networks is driven in part by the success of neural networks in visual object recognition. Artificial neural networks are also good at analysing large sets of unlabeled, often high-dimensional data – where it may be difficult to determine a priori which questions are most relevant and rewarding to ask.

The tools for machine learning with neural networks were developed long ago, most of them during the second half of the last century. In 1943, McCulloch and Pitts \cite{Pitts} analysed how networks of neurons can process information. Using an abstract model for a neuron, they demonstrated how such units can be coupled together to represent logical functions (Fig. 1).

\begin{figure}
    \centering
    \includegraphics[width=0.45\textwidth]{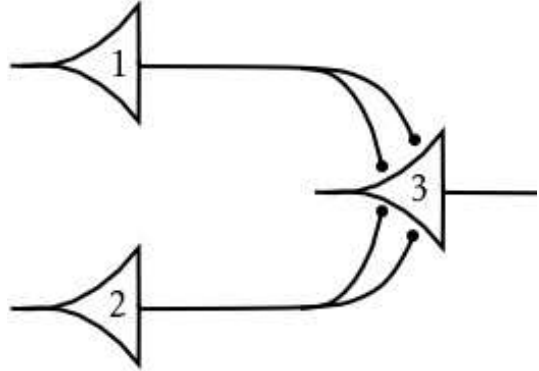}
    \caption{Logical OR function represented by three neurons. Neuron 3 fires actively if at least one of the neurons 1 and 2 is active. After Fig. 1b by McCulloch and Pitts \cite{Pitts}}
    
\end{figure}

\subsection{Applications in high-energy physics}

Artificial neural networks are the machine learning technique best known in the high-energy physics community where important physics results have been extracted using this method in the last decade. In terms of types of ANNs, the vast majority of applications in high-energy physics are based on feed-forward multi-layer ANN with back-propagation.

In terms of types of applications, ANN were used for both online triggers and offline data analysis. For offline data analysis ANNs were used or tested for a variety of tasks such as track and vertex reconstruction, particle identification and discrimination, calorimeter energy estimation and jet tagging. The first application from which a physics results was extracted with an ANN was for the decay of the Z boson \cite{opal}. A feed-forward network was used to discriminate the decay of the Z boson into c, b or s quarks and the results used further to determine the decay probability of Z into the corresponding states. Here, in this article, we intent to study the feed-forward multi-layer ANN with back-propagation which has a vast majority of applications in high-energy physics.

The process in consideration is the Higgs to to di-tau where one tau decays to leptonically  and while the other hadronically giving the final state as l, $\nu$, jets. The Feynman diagram of the tau decay process is shown in Fig. 2. The invariant mass plot of the final is shown in Fig. 3 which shown that there is large overlap of both signal and background. Thus the usual method of cut and cound is no longer applicable for the 5$\sigma$ discovery. Thus Machine Learning Technique is needed  to maximize the signal over background for the processes for which there is no distinct separation.

\begin{figure*}[htb]
\begin{minipage}[t]{.5\textwidth}
\centering
\includegraphics[width=\textwidth]{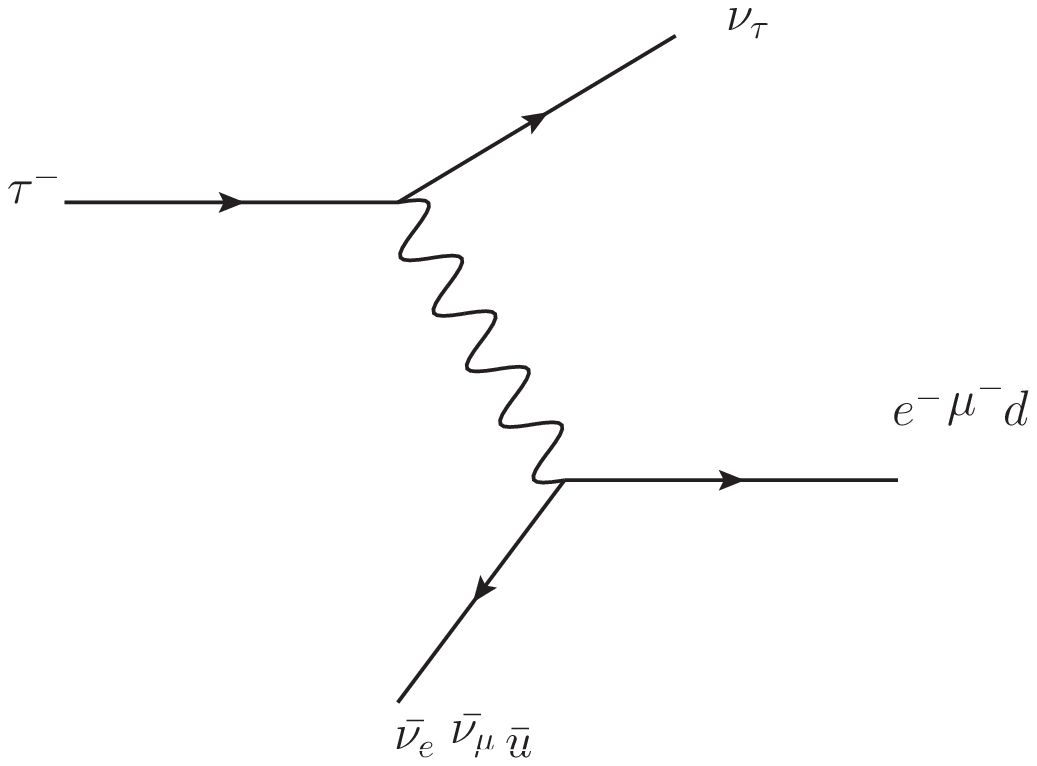}
\caption{Feynman diagram showing one of the tau lepton decays.}
\end{minipage}
\hfill
\begin{minipage}[t]{.5\textwidth}
\centering
\includegraphics[width=1.0\textwidth]{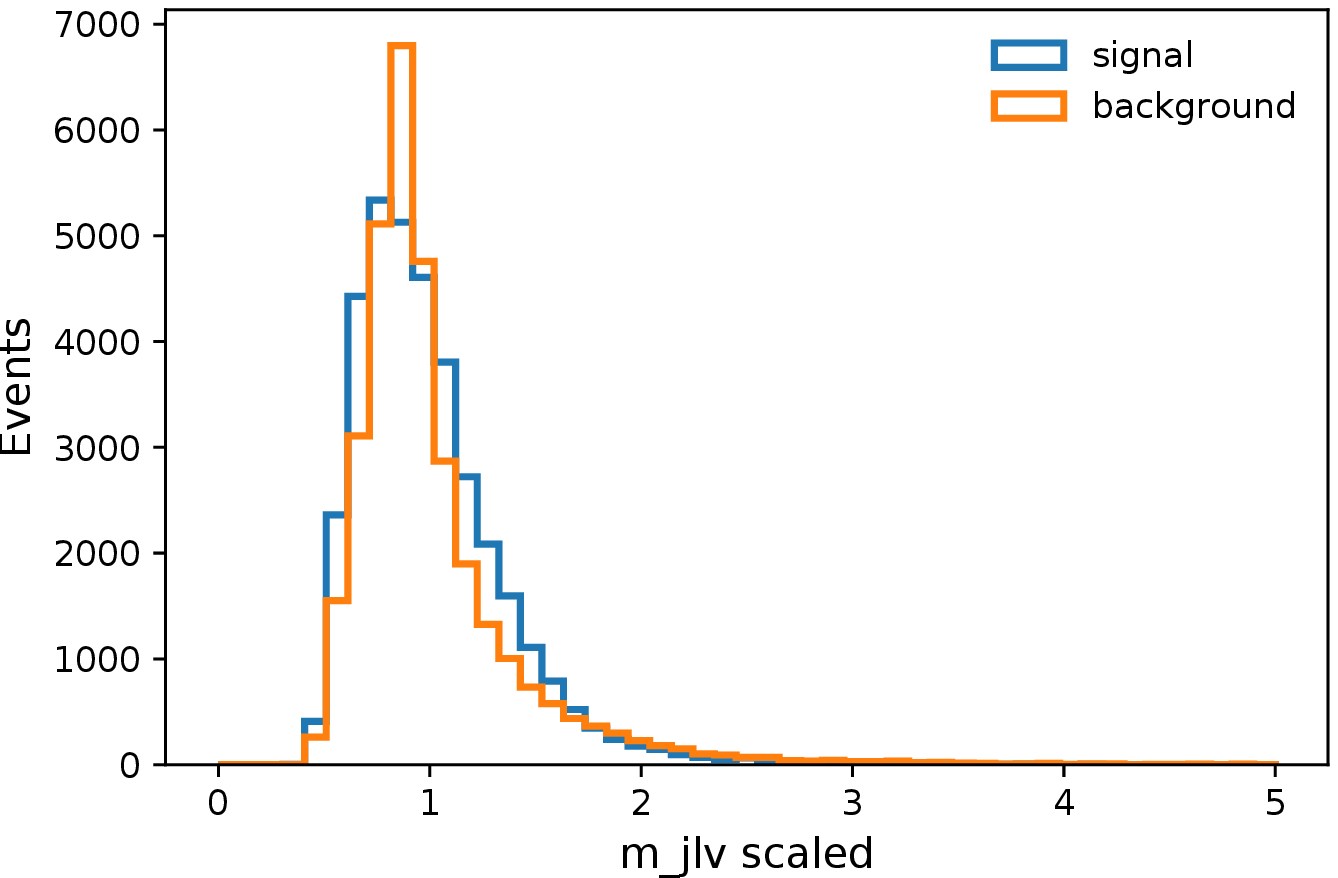}
\caption{Scaled invariant mass of the final states of the $H\rightarrow \tau \tau$ process}
\end{minipage}
\end{figure*}

\section{Multi-layer ANNs}

\begin{figure}
   \centering
    \includegraphics[width=0.6\textwidth]{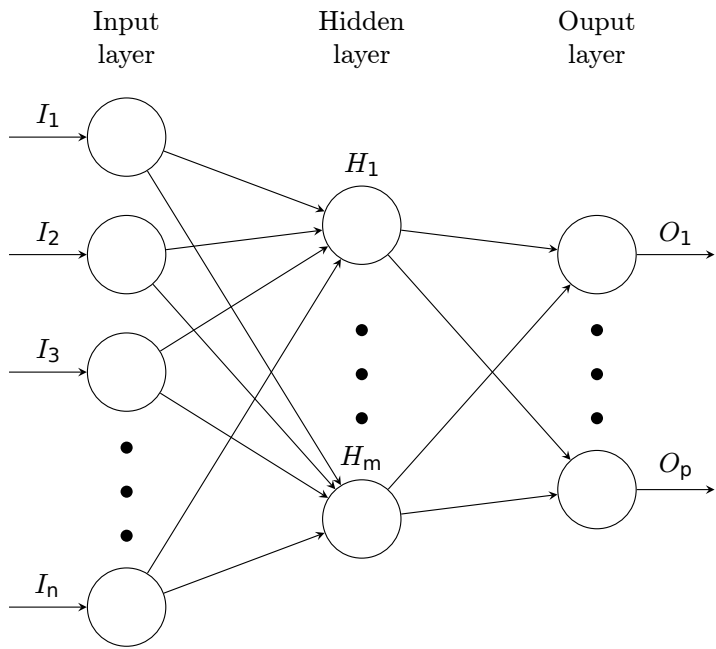}
    \caption{A multi-layer ANN with input, hidden and output layers }
\end{figure}

A single AN is quite limited in its functionality as it can produce linearly separable functions only. For more complex functions a network of interconnected ANs are necessary. One simple way to organise the ANs is in layers resulting a class of ANN called multi-layer ANNs. An example of a muti-layer ANN is shown in Fig. 4. It contains a layer of n input neurons $I_{i}$ (i = 1, ..., n), a hidden-layer of m neurons $H_{j}$ with the activation function $f_{Hj}$ (j = 1, ..., m) and an output layer of p neurons $O_{k}$ with the activation function $f_{Ok}$ (k = 1, ..., p). The activation functions could be different for different layers. The connections between the neurons are weighted with different values. $v_{ji}$ are the weights between the input layer and the hidden layer, and $w_{kj}$ are the weights between the hidden layer and the output layer. Using these weights, the network propagates the external signal through the layers producing the output signal which is of the form
\begin{center}
\begin{equation}
\fl    O_{k} = f_{Ok}(net_{Ok}) = f_{Ok}\left( \sum_{j=1}^{m+1} w_{kj}f_{Hj}(net_{Hj}) \right) = f_{Ok}\left( \sum_{j=1}^{m+1} w_{kj}f_{Hj}\left( \sum_{i=1}^{n+1} v_{ji}I_{i} \right) \right)
\end{equation}
\end{center}

This type of ANN, which simply propagates the input through all the layers, is called a feed-forward multi-layer ANN. The example discussed here contains only one hidden layer. The network architecture can be extended to contain as many hidden layers as necessary.

\subsection{ANN learning}
The learning process for an ANN is the process through which the weights of the network are determined. This is achieved by adjusting the weights until certain criteria are satisfied. There are three main types of learning.
\begin{enumerate}
    \item supervised learning
    \item unsupervised learning
    \item reinforcement learning
\end{enumerate}
Here, in this article, we discuss the supervised learning which is used in the vast majority of the high-energy physics applications. Here,the ANN is presented with a training dataset which contains the input vectors and a target associated with each input vector. The target is the desired output. The weights of the ANN are adjusted iteratively such that the difference between the actual output of the ANN and the target is minimized.
The most common supervised learning method is based on the gradient descent learning rule. The method optimises the network weights such that a certain objective function E is minimised by calculating the gradient of E in the weight space and moving the weight vector along the negative gradient. The binary crossentropy (BCE) is used for training. The BCE calculates the loss by computing the following average:
\begin{equation}
    \fl Loss = - \frac{1}{N} \Sigma_{i}^{N} O_{i}\cdot log \hat{O_{i}} + (1-O_{i})log(1-\hat{O_{i}})
\end{equation} 
where $\hat{O_{i}}$ is the i-th scalar value in the output, $O_{i}$ is the corresponding target value, and N is the number of scalar values in the model output.
For each iteration (usually called epoch), the gradient descent weight optimisation contains two phases:
\begin{itemize}
    \item feed-forward pass in which the output of the network is calculated with the current value of the weights, activation function and bias
    \item backward propagation in which the errors of the output signal are propagated back from the output layer towards the input layer and the weights are adjusted as a function of the back-propagated errors.
\end{itemize}
To summarise, the supervised learning process implies the following steps:
\begin{enumerate}
    \item initialisation of the weights
     \item initialisation of the loss function
     \item for each training pattern
     \begin{enumerate}
         \item calculate $H_{jp}$ and $O_{kp}$ (feed-forward phase)
         \item calculate the output error and the hidden layer error
         \item adjust the weights $w_{kj}$ and $v_{ji}$ (back-propagation phase)
         \item update of the loss function
     \end{enumerate}
     \item test the stopping criteria; if this is not met then the process continues from the step (ii)
     
\end{enumerate}
As stopping criteria, common choices are a maximum number of epochs, a minimum value of the error function evaluated for the training data set, and the over-fitting point. The initialisation of the weights, the learning rate and the momentum are very important for the convergence and the efficiency of the learning process.

\subsection{Scaling}
Machine learning models may have difficulty converging before the maximum number of iterations allowed if the data aren’t normalized. We must apply the same scaling to the test set for meaningful results. There are a lot of different methods for normalization of data such as  using python module numpy to scale our training sample and test samples where all numerical attributes are scaled to have a mean of 0 and a standard deviation of 1. But here we use StandardScaler for normalization of data before they are fed for machine learning technique.

\subsection{Regularisation}
If one uses the backpropagation algorithm to find a solution to this problem, one may find that the weights continue to grow during training. This can be problematic because it may imply that the local fields become so large that the activation functions saturate. Deeper networks have more neurons, so the problem of overfitting  tends to be more severe for deeper networks. Regularisation schemes limit the tendency to overfit. A number of regularisation schemes such as cross validation, weight decay, pruning, drop out, batch normalisation etc. have proved useful for deep networks. Here, in our analysis we have applied drop out and batch normalization. Batch normalization mitigate internal covariate shift through normalization of the layers' inputs by re-centering and re-scaling.

\subsection{Choice of activation function}

%%%%%%% Figure for the activation function

\begin{figure*}[htb]
\begin{minipage}[t]{.49\textwidth}
\centering
\includegraphics[width=\textwidth]{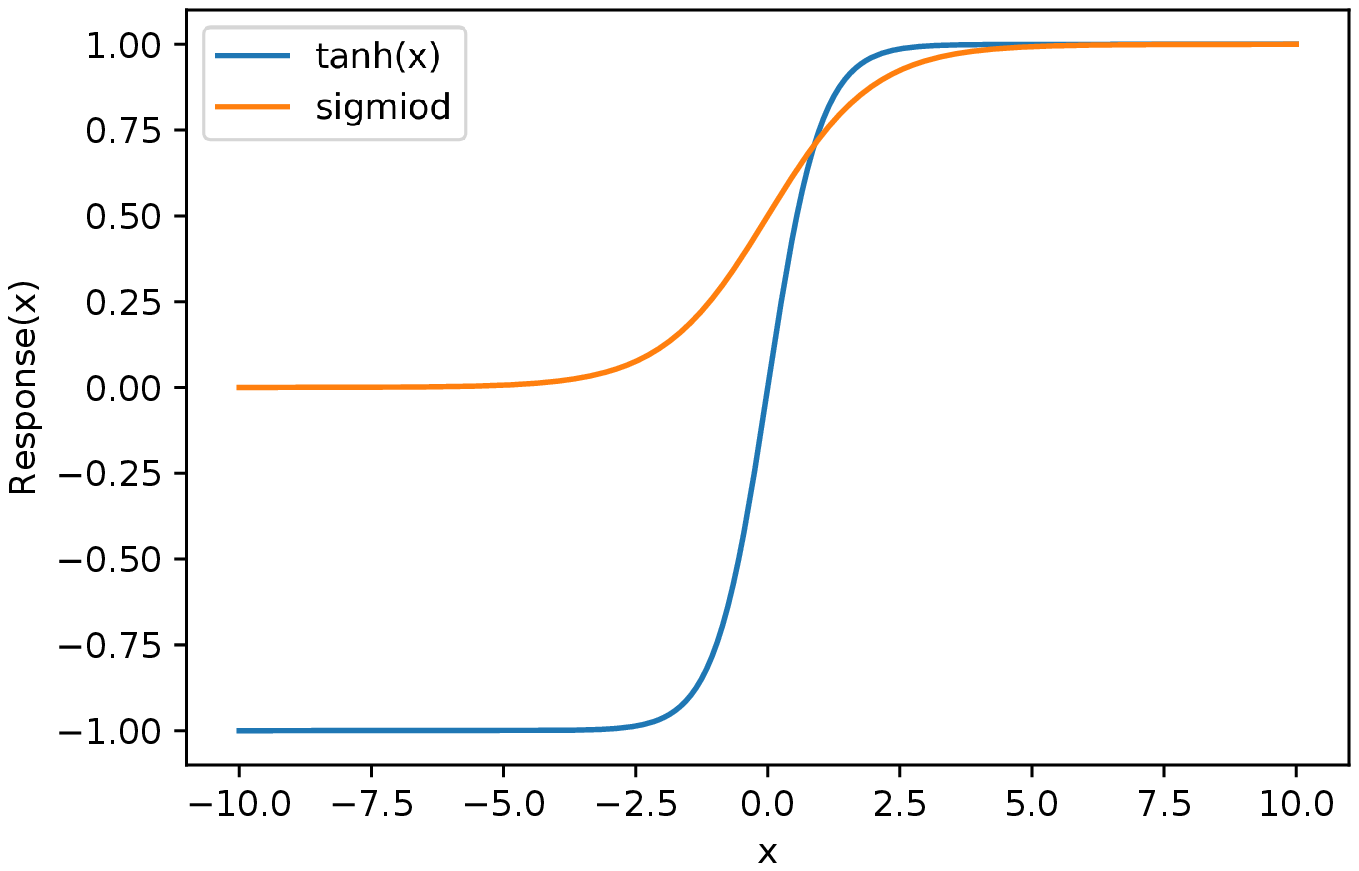}
\caption{Response of tanh(x) vs Sigmoid function}
\end{minipage}
\hfill
\begin{minipage}[t]{.49\textwidth}
\centering
\includegraphics[width=\textwidth]{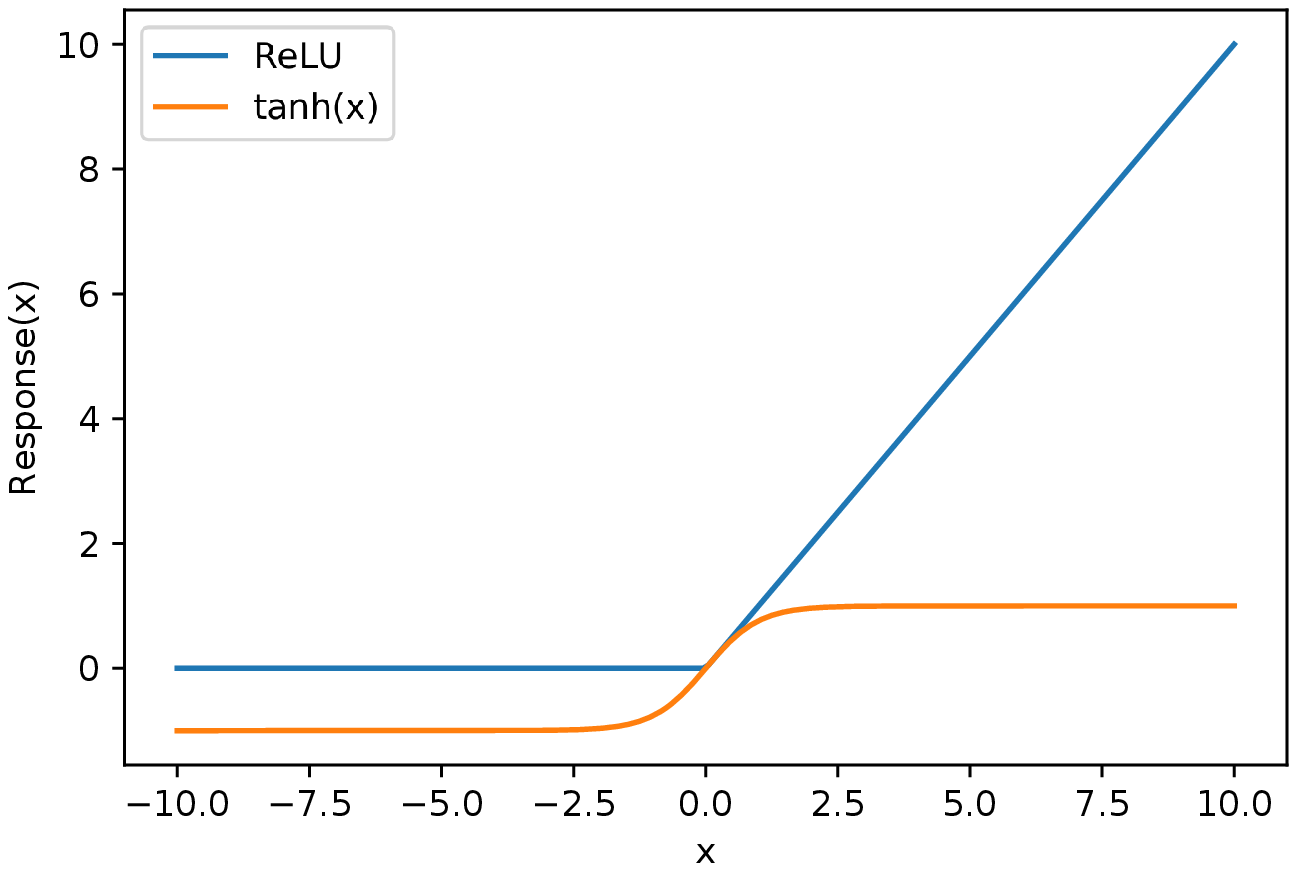}
\caption{Response of tanh(x) vs ReLU function}
\end{minipage}
\end{figure*}

The ideal activation function would be a signum function  but the jump in the signum function at x=0 may cause large fluctuations in the activity levels of a network of neurons, caused by infinitesimal changes of the local fields across x = 0. To dampen this effect, one allows the neuron to respond continuously to its inputs. The possible activation function that can be used in Neural Networks are linear functions, sigmiod functions, tanh(x), ReLU functions. From the response curve of the dfferent functions shown in Fig. 5 \& Fig. 6, it is clear that ReLU is the most preferred and then tanh(x) over Sigmoid function. Here, we present the results where the input layers are fed into the Network using eLU activation function. The hidden layers are fed with ReLU activation function. The sigmoid activation function is used for the output layer.

\section{Results}
\subsection{Construct an Artificial Neural Network:}
\begin{figure*}[htb]
\begin{minipage}[t]{.49\textwidth}
\centering
\includegraphics[width=\textwidth]{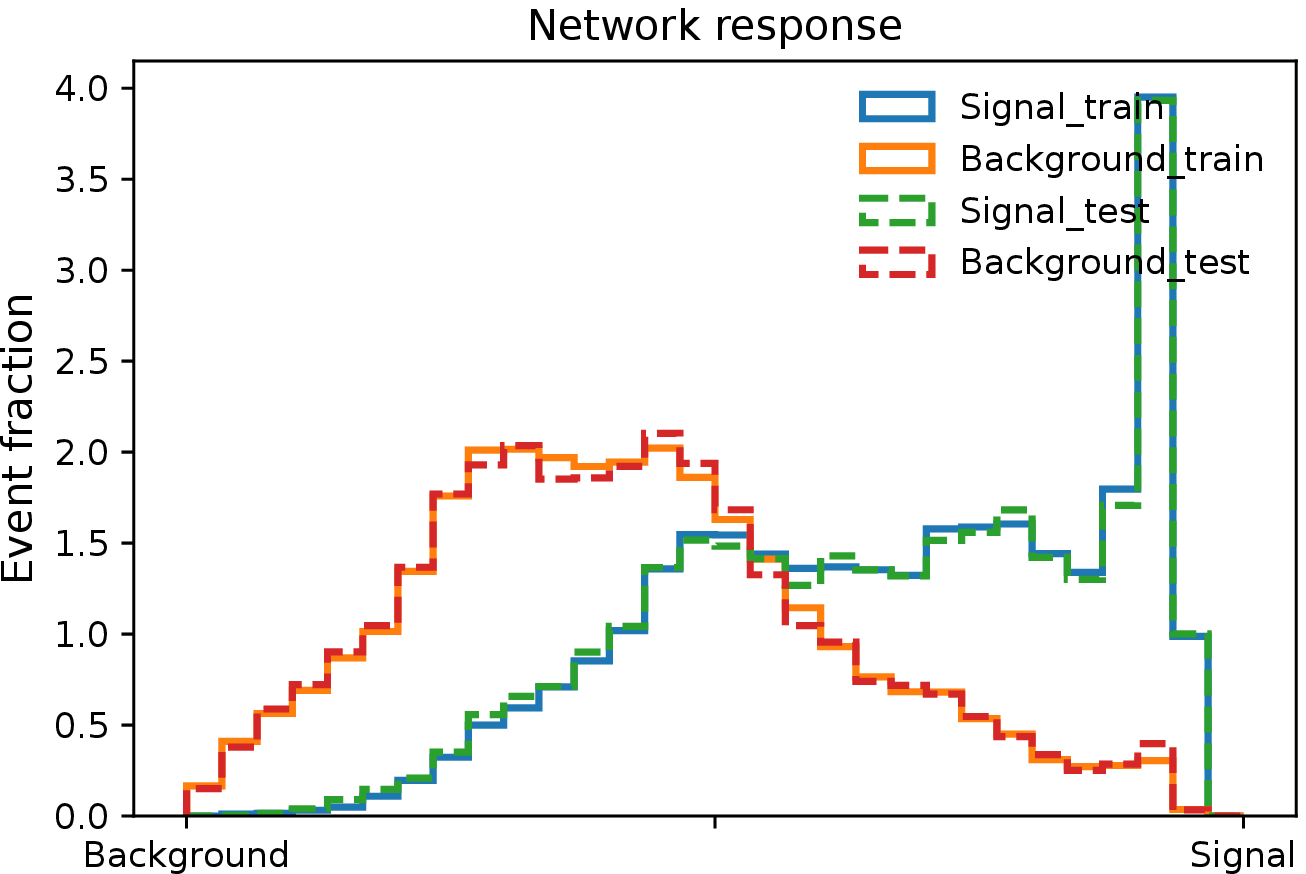}
\caption{Output of the ANNs for the training and test samples}
\end{minipage}
\hfill
\begin{minipage}[t]{.49\textwidth}
\centering
\includegraphics[width=\textwidth]{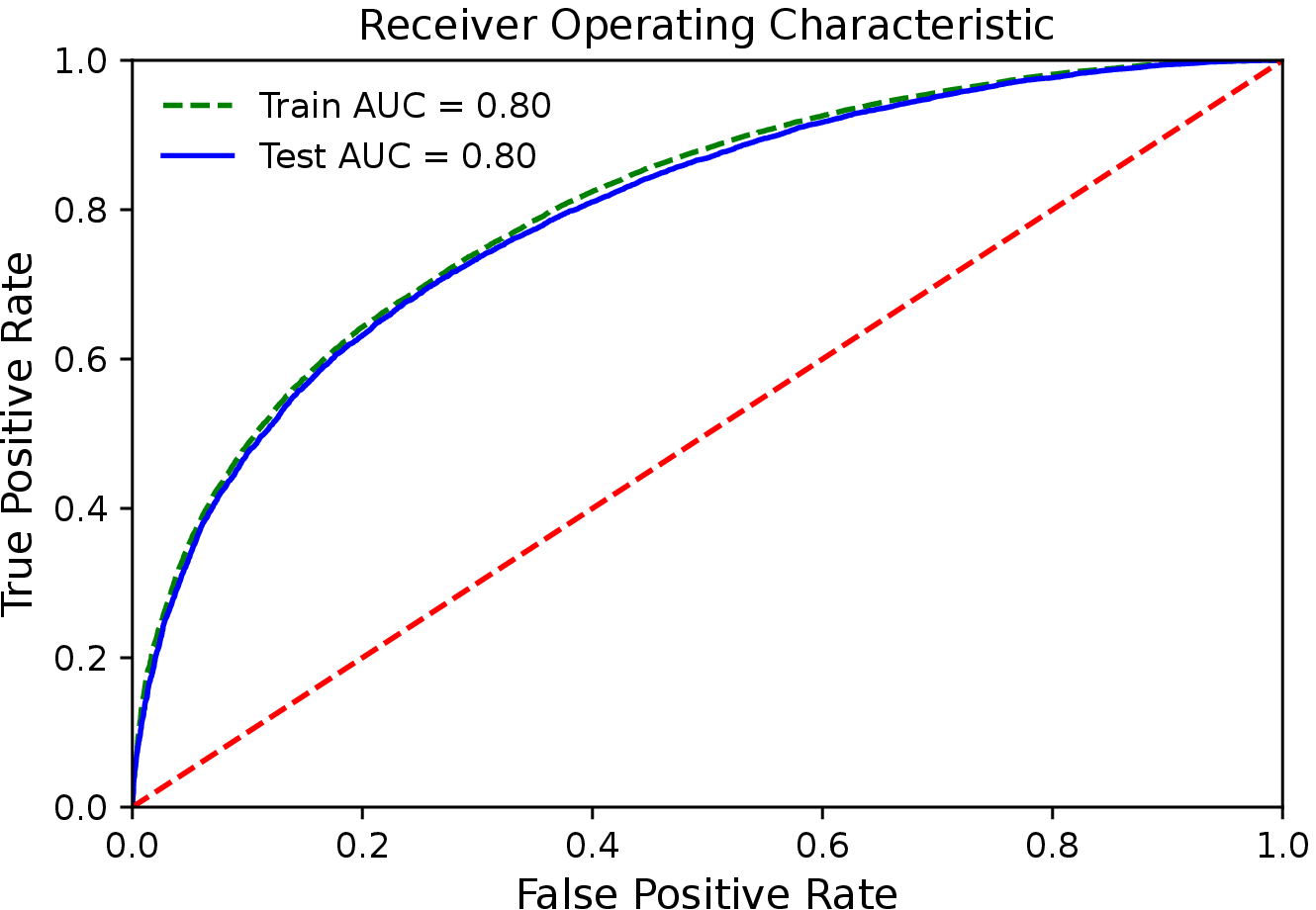}
\caption{ROC curve for for train and test data samples}
\end{minipage}
\end{figure*}

\begin{figure*}[ht]
\begin{minipage}[t]{.49\textwidth}
\centering
\includegraphics[width=\textwidth]{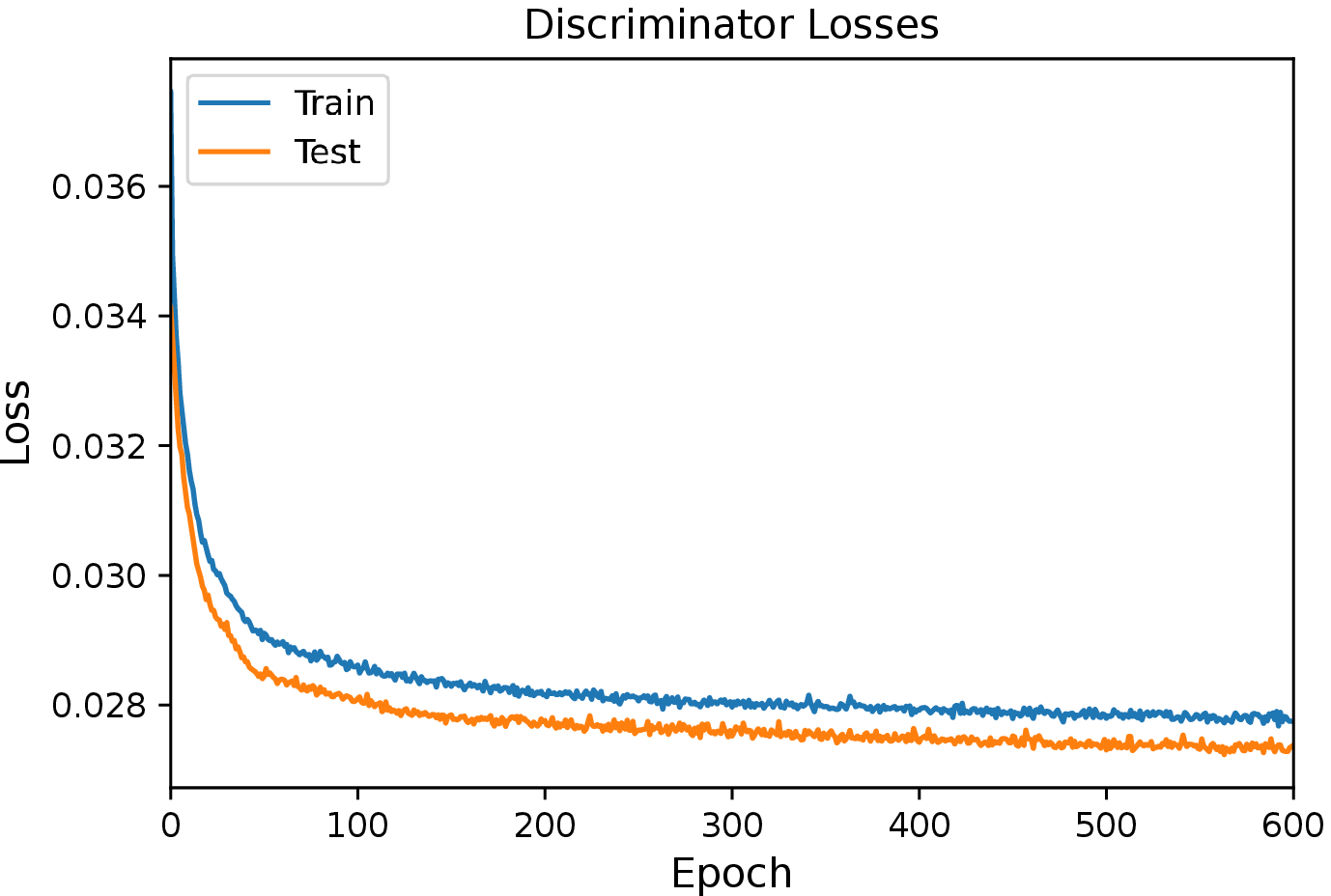}
\caption{losses of the Neural Network configuration with number of iterations for train and test samples}
\end{minipage}
\hfill
\begin{minipage}[t]{.49\textwidth}
\centering
\includegraphics[width=\textwidth]{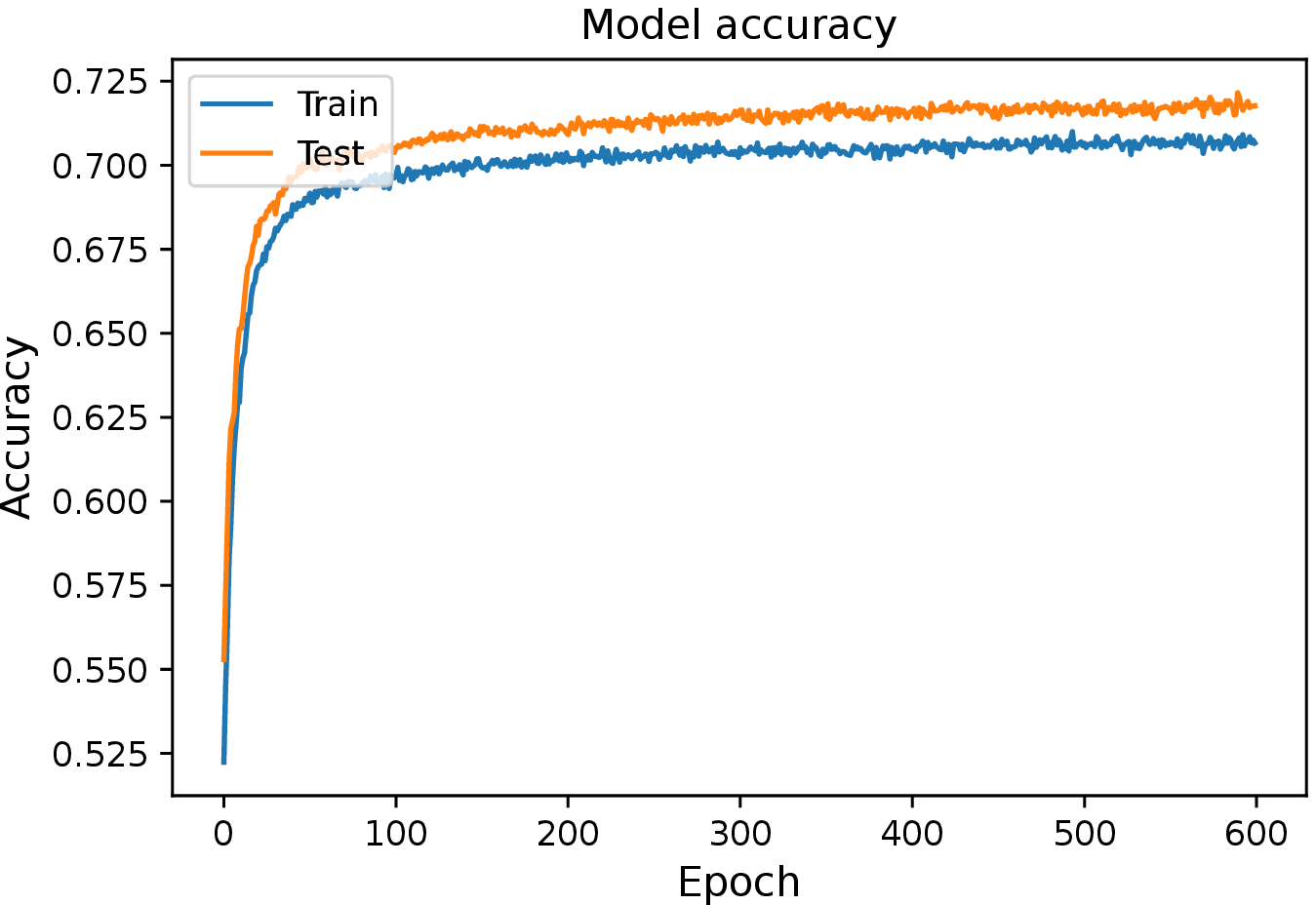}
\caption{Accuracy of the Neural Network model for the train and test samples}
\end{minipage}
\end{figure*}

A multi-layer Neural Network as a classifier to distinguish signal from background for $H \rightarrow \tau \tau$ is constructed in python using Tensorflow\cite{tensorflow}. The weight associated are also generated using random number generator in python. A Neural Network model with 20 nodes, 3 discriminator layers with adam optimizer and binary crossentropy used for optimization and loss calculation. The small learning rate avoids overshooting. Fig. 7 shows the neural network output for the signal and background distributions for the training and testing samples. Fig. 8 shows the  ROC curve of the neural network output. From the area under the ROC curve, it is clear that the efficiency of our neural network training is good and there is no over-fitting of data. Fig. 9 \& Fig. 10 show the loss  and the accuracy of the neural network setup with varying the number of the epoches. The loss decreases with the increase in the number of epoches while the model accuracy increases with the increase in the number of epoches. The cut around 0.5 on the Network response will maximize the signal over background.

\begin{figure}
    \centering
    \includegraphics[width=0.5\textwidth]{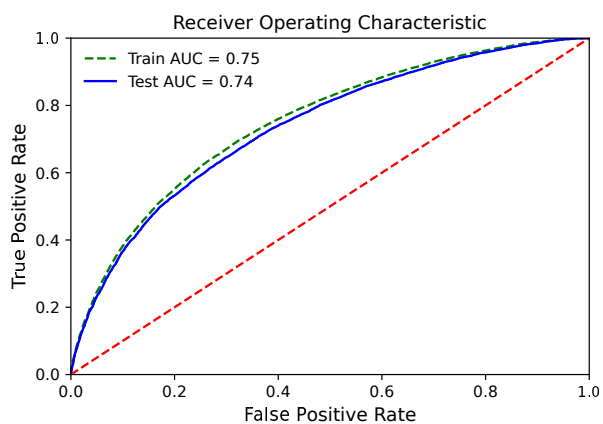}
    \caption{ROC curve without the use of regularization for the same hyperspace parameters.}
\end{figure}

\begin{figure*}[htb]
\begin{minipage}[t]{.5\textwidth}
\centering
\includegraphics[width=\textwidth]{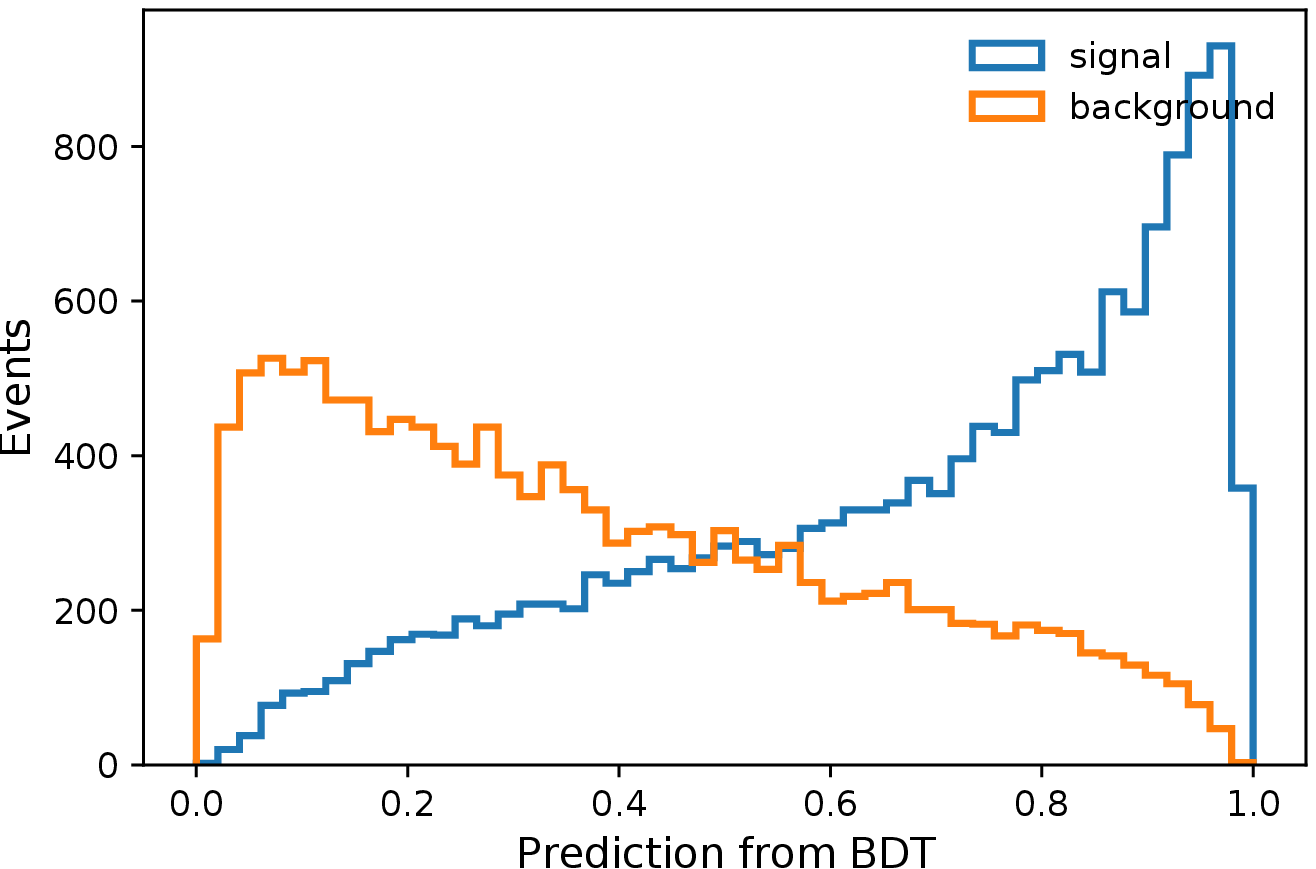}
\caption{BDT score for the signal and background for the test sample}
\end{minipage}
\hfill
\begin{minipage}[t]{.5\textwidth}
\centering
\includegraphics[width=\textwidth]{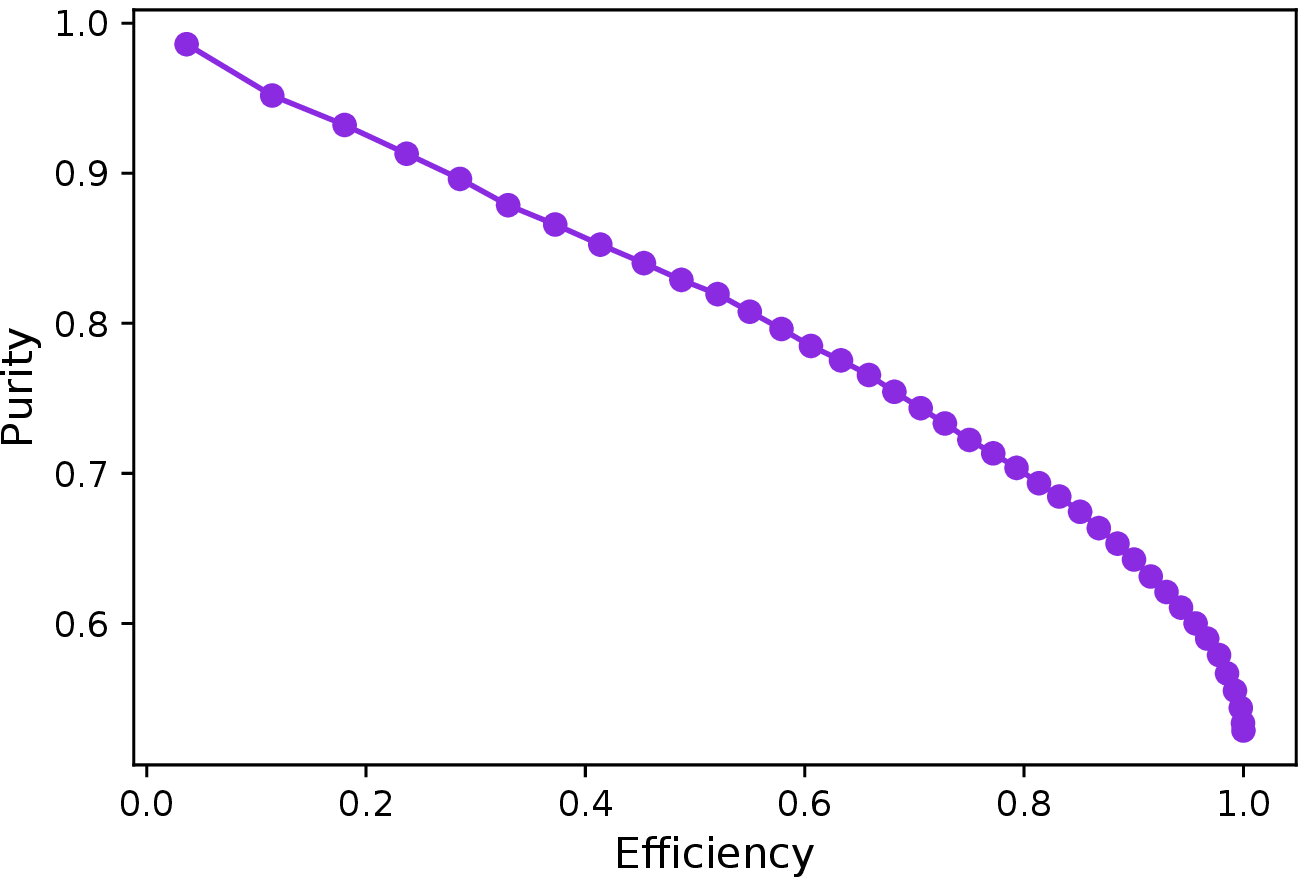}
\caption{Efficiency of the BDT Network}
\end{minipage}
\end{figure*}

\begin{figure}
    \centering
    \includegraphics[width=0.86\textwidth]{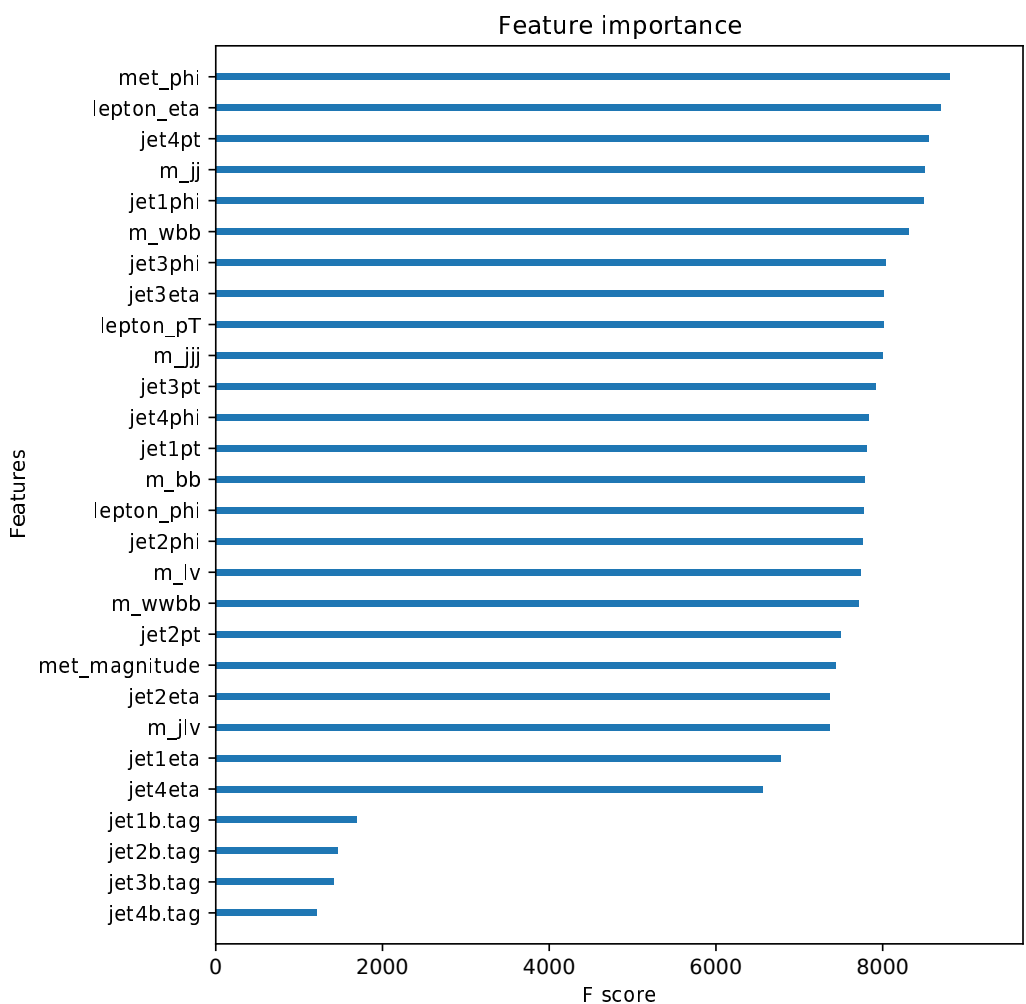}
    \caption{Ranking of all the input features using the machine learning algorithm XGBoost based on BDT prediction.}
\end{figure}

\subsection{Regularization}
The regularization is performed using the method discussed in Section 2.3. We observed the efficiency of the ANN increases with use of regularization. Fig. 11 shows the ROC curve for the Neural Network for the same hyperspace parameters without use of regularization.  The area under the ROC curve without the use of the regularization decreases. This also has large effect on the response of the neural network.

\section{Ranking of the parameters}
Ranking is a subset of supervised machine learning that put the features in the order that maximizes the signal efficiency. For that we use a powerful machine learning algorithm named XGBoost\cite{xgboost}. It implements gradient boosting algorithm and ranks the input features based on the score. The BDT prediction for our test sample is shown in Fig. 12 and the purity of signal in Fig. 13. Here the signal and the background is exactly 50\% but well-separated. The ranking of all the input features is shown in Fig. 14. From Fig. 14, missing energy phi is the feature that has maximum significance where as jet4b tag feature has the least.

\section{Conclusion}
A multi-layer Neural Network as a classifier to distinguish signal from background was constructed.  It was found that for higher number of epoch and smaller learning rate, the Network performs better while the regularization technique such as batch normalization increase the network efficiency for the same hyperspace parameters and also reduces the over-fitting of data. It was found that due to back propagation, the loss continue to grow due to which the local field become so large that the activation function saturate.

Based on the BDT response and the efficiency of the BDT network, the ranking of the input features were predicted.  It was found that met$\_$phi had the highest significance while jet4b,tag had the least significance. This study can be a basis for understanding the use of Machine Learning Technique in high energy physics.

\end{document}